\documentclass[aps,reprint,amsmath,amssymb,groupaddress,superscriptaddress,prl]{revtex4-2}

\usepackage{xcolor}

\usepackage[utf8]{inputenc}
\usepackage{amsmath}
\usepackage{amsfonts}
\usepackage{amssymb}
\usepackage{bm}
\usepackage{color}
\usepackage{graphicx}
\usepackage{soul}
\usepackage{hyperref}
\usepackage{CJK}

\usepackage{mathtools}
\usepackage{lipsum}

\newcommand{\sst}{\mathrm{st}}
\newcommand{\res}{\mathrm{res}}
\newcommand{\out}{\mathrm{out}}
\newcommand{\tot}{\mathrm{tot}}
\newcommand{\sys}{\mathrm{sys}}

\begin{document}

\begin{CJK*}{UTF8}{gbsn}
\title{Thermodynamic bounds on time-reversal asymmetry}
\author{Shiling Liang (梁师翎)}
\affiliation{Institute of Physics, School of Basic Sciences, \'Ecole Polytechnique F\'ed\'erale de Lausanne (EPFL), 1015 Lausanne, Switzerland}
\affiliation{Biological Complexity Unit, Okinawa Institute of Science and Technology Graduate University, Onna, Okinawa 904-0495, Japan}
\author{Simone Pigolotti}
\affiliation{Biological Complexity Unit, Okinawa Institute of Science and Technology Graduate University, Onna, Okinawa 904-0495, Japan}

\begin{abstract}
Quantifying irreversibility of a system using finite information constitutes a major challenge in stochastic thermodynamics. We introduce an observable that measures the time-reversal asymmetry between two states after a given time lag.  Our central result is a bound on the time-reversal asymmetry in terms of the total cycle affinity driving the system out of equilibrium. This result leads to further thermodynamic bounds on the asymmetry of directed fluxes; on the asymmetry of finite-time cross-correlations; and on the cycle affinity of coarse-grained dynamics.
\end{abstract}
\maketitle
\end{CJK*}

Non-equilibrium systems are characterized by their irreversible dynamics.  This irreversibility is generated by dissipative forces, and thus implies a thermodynamic cost \cite{seifert2019stochastic,yang2021physical}. In stochastic thermodynamics \cite{peliti2021stochastic}, irreversibility is encoded in statistical properties of stochastic trajectories \cite{seifert2005entropy}. In practice, measuring probabilities of whole trajectories is often hard, and one is forced to estimate irreversibility from incomplete statistical information. For example, fluctuations of currents and first-passage time can be used to set bounds on the dissipation rate through thermodynamic uncertainty relation \cite{barato2015thermodynamic,li2019quantifying,pietzonka2016universal,gingrich2017fundamental,pal2021thermodynamic}. Responses to perturbations can also reveal non-equilibrium properties \cite{owen2020universal}. Often, only a few states of a physical system may be ``visible'', i.e. experimentally observable, and the temporal resolution of measurement may be also finite. Inferring thermodynamic costs from such partial observations remains an active field of investigation \cite{martinez2019inferring,harunari2022learn,van2022thermodynamic,van2023time,cisneros2023dissipative}. 

\begin{figure}[htb]
    \centering
    \includegraphics[width=1\columnwidth]{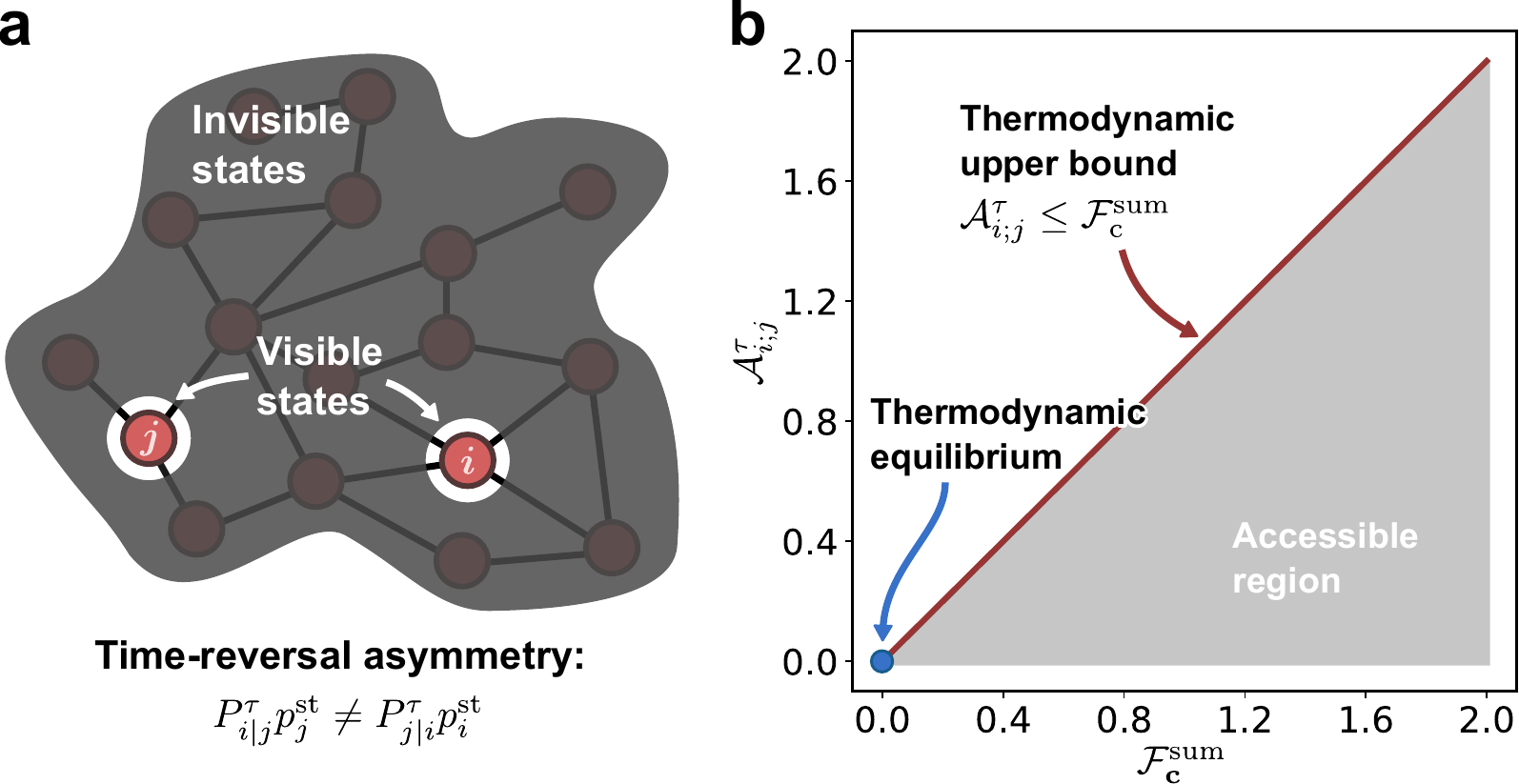}
    \caption{(a) A Markov network with only two visible states. The time-reversal symmetry can be broken by driving the system out of equilibrium. (b) Thermodynamic bound on the time-reversal asymmetry $\mathcal{A}_{i;j}^\tau$, as defined in Eq.~\eqref{eq:relativejoint}, Here, $\mathcal{F}_{\mathbf{c}}^{\mathrm{sum}}=\sum_{\mathbf{c}}|\mathcal{F}_{\mathbf{c}}|$ is the sum of all cycle affinities in the network.}\label{fig:time-reversal asymmetry}
\end{figure}

In this Letter, we propose and study the following measure of time-reversal asymmetry between two states:
\begin{equation}\label{eq:relativejoint}
  \mathcal{A}_{i;j}^\tau \equiv  \left|\ln \left(\frac{P_{i|j}^\tau p_j^{\sst}}{P_{j|i}^\tau p_i^{\sst}}\right)\right| ,
\end{equation}
where $p_i^{\sst}$ is the probability of state $i$ at steady state and $P_{i|j}^\tau$ is the propagator, defined as the probability of finding the system in state $i$ after a time lag $\tau$, given that it initially was in state $j$, see Fig.~\ref{fig:time-reversal asymmetry}a.  The time-reversal asymmetry $\mathcal{A}_{i;j}^{\tau}$ must be equal to zero for any choice of $i$, $j$, and $\tau$ at equilibrium, where forward and backward trajectories have equal probability and thus time-reversal symmetry is preserved. 

The freedom in choosing the states $i$, $j$, and the time lag $\tau$ allows us for using $\mathcal{A}_{i;j}^\tau$ to probe how the non-equilibrium nature of a system affects its different states at different timescales. In the short time limit, only the direct edge between state $i$ and $j$ contributes the propagator, so that the time-reversal asymmetry reduces to the absolute log ratio of directed fluxes, $\lim_{\tau\to0} \mathcal{A}_{i;j}^{\tau}=|\ln[W_{ij}p^\sst_j/(W_{ji}p^\sst_i)]|$. In contrast, in the long time limit, propagators lose memory of the initial state, so that $\lim_{\tau\to\infty} P_{i|j}^{\tau}=p_i^{\sst}$ and $\lim_{\tau\to\infty} \mathcal{A}_{i;j}^{\tau}=0$.

Our central result is that, out of equilibrium, the time-reversal asymmetry $\mathcal{A}_{i;j}^{\tau}$ is bounded from above by the sum of affinities driving all the cycles in the network, irrespectively of the choice of $i$, $j$, and $\tau$, see Fig.~\ref{fig:time-reversal asymmetry}b. To prove our result, we combine trajectory thermodynamics with a graph-theoretic description of Markov networks \cite{schnakenberg1976network,hill1989free,maes2013heat,liang2022universal}. We conclude with three applications to other observables: the asymmetry of directed fluxes; the asymmetry of finite-time cross-correlations; and the cycle affinity of temporal-coarse-grained dynamics.

\medskip

{\em Setup.} We consider a discrete-state system whose dynamics is described by a master equation
\begin{equation}\label{eq:mastereq}
    \frac{d}{dt}\mathbf{p} = W\mathbf{p} \ ,
\end{equation}
where $\mathbf{p} = \mathbf{p}(t) = (p_1(t),\cdots,p_n(t))$ is the probability distribution on a set of $n$ states. The off-diagonal elements $W_{ij}$ of the matrix $W$ represent the transitions rates from state $j$ to state $i$, whereas $W_{ii}=-\sum_{j\neq i}W_{ji}$. Thermodynamic properties are encoded in the local detailed balance (LDB) condition
\begin{equation}\label{eq:ldb}
W_{ij}/W_{ji}=e^{s^\res_{e_{ij}}},
\end{equation}
which associates the transition rates between states $i$ and $j$ with the entropy $s^\res_{e_{ij}}$ released into the thermal reservoir when going from state $j$ to $i$ through the edge $e_{ij}$, see Refs.~\cite{maes2021local,peliti2021stochastic}. Here and in the following, we set the temperature $T=1$ and the Boltzmann constant $k_\mathrm{B}=1$.

The steady-state probability distribution is the eigenvector of $W$ associated with the zero eigenvalue, $W\mathbf{p}^\sst=0$. Such eigenvalue is unique according to Perron-Frobenius theorem.

The formal solution of Eq.~\eqref{eq:mastereq} is
\begin{equation}\label{eq:finite_time_evol}
\mathbf{p}(\tau) = e^{W \tau}\mathbf{p}(0).
\end{equation}
The propagator $P_{i|j}^{\tau}$ is the $ij$ element of the matrix $e^{W\tau}$. At steady state, the joint probability of two states separated by a lag time $\tau$ is
\begin{equation}\label{eq:joint prob}
  P_{i;j}^{\tau}\equiv  P_{i|j}^{\tau} p_j^{\sst}.
\end{equation}

We shall derive thermodynamic bounds to the relative joint probabilities based on graph-theoretic concepts, that we briefly introduce in the following. A walk ($\mathbf{w}_{ij}$) is a sequence of directed edges ($e$) which join a certain sequence of states ($v$) from state $j$ to state $i$. A path ($\mathbf{r}_{ij}$) is a walk from $j$ to $i$ with neither repeated edges nor vertices. A cycle ($\mathbf{c}$) is a  closed path. The entropy production released into the heat reservoir along a given path from state $j$ to state $i$ is the sum of entropy released for all the edges belonging to this path:
\begin{equation}
    s^\res_{\mathbf{r}_{ij}} =\ln\left[ \prod_{e\in \mathbf{r}_{ij}} \frac{W_e^+}{W_{e}^-}\right] = \sum_{e\in \mathbf{r}_{ij}} s^\res_e.
\end{equation}
where $W_e^{+/-}$ is the forward(backward) rate along the edge $e$, $s^\res_e$ is the entropy released into the reservoir along the same edge, and the second equality follows from Eq.~\eqref{eq:ldb}. At equilibrium, the entropy $s^\res_{\mathbf{r}_{ij}}$ is independent of the chosen path, and thus the cycle affinity $\mathcal{F}_{\mathbf{c}}=\sum_{e\in \mathbf{c}} s^\res_e$ vanishes for any cycle \cite{maes2021local}.

\begin{figure}[t]
    \centering
    \includegraphics[width=1\columnwidth]{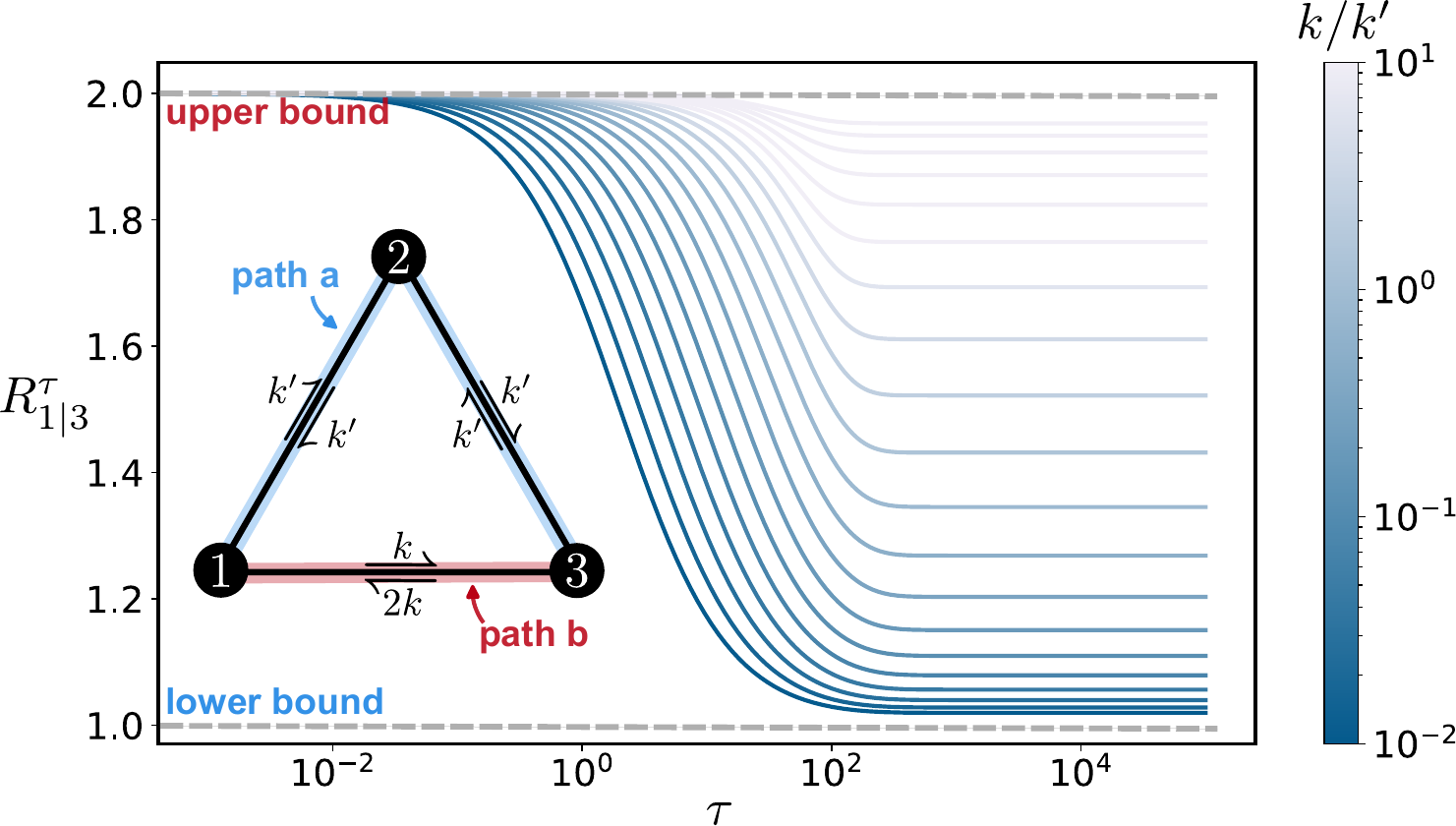}
    \caption{The relative propagator $R_{1|3}^{\tau}$ as a function of the lag time $\tau$ for different kinetic rates. In the small $\tau$ limit, $R_{1|3}^{\tau}$ is determined by the path b and approaches its upper bound. In the large $\tau$ limit, $R_{1|3}^{\tau}$ approaches the lower bound for small $k/k'$.}
    \label{fig:condi_asymmetry}
\end{figure}
{\em Bound on relative propagators.} Our first main result is that the relative propagator $R_{i|j}^{\tau}\equiv  P_{i|j}^\tau/P_{j|i}^\tau $  is bounded by
\begin{equation}
    \label{eq:conditional}
    e^{s_{\mathbf{r}_{ij}}^{\res,\min}}\leq R_{i|j}^{\tau} \leq e^{s_{\mathbf{r}_{ij}}^{\res,\max}}.
\end{equation}
Here, $s_{\mathbf{r}_{ij}}^{\res,\max/\min}$ is the maximum/minimum entropy released into the reservoir among all possible paths $\mathbf{r}_{ij}$. In a three-state network, we find that the relative propagator $R_{1|3}^{\tau}$ ranges between the upper and lower bounds as $\tau$ increases, see Fig.~\ref{fig:condi_asymmetry}. Equation~\eqref{eq:conditional} can be seen as a finite-time LDB condition (see Eq.~\eqref{eq:ldb}), since, if two states $i$ and $j$ are connected and in the short time limit, the relative propagator reduces to the ratio of transition rates:
\begin{equation}
\lim_{\tau\rightarrow 0} R_{i|j}^{\tau}=W_{ij}/W_{ji} .
\end{equation}

To prove Eq.~\eqref{eq:conditional}, we represent the propagator from $j$ to $i$ as a sum of contributions from all possible walks from $j$ to $i$:
\begin{equation}
    P_{i|j}^{\tau}= \sum_{\mathbf{w}_{ij}}P(\mathbf{w}_{ij};\tau),
\end{equation}
where $P(\mathbf{w}_{ij};\tau)$ is the probability of all the trajectories associated with the walk $\mathbf{w}_{ij}$ in a time $\tau$. Such probability is expressed by
\begin{equation}
    P(\mathbf{w}_{ij};\tau) = P_{\mathbf{w}_{ij}}^\tau \prod_{e\in\mathbf{w}_{ij}}W_e^+, \label{eq:walktau}
\end{equation}
where $P_{\mathbf{w}_{ij}}^\tau=\int \prod_{\nu\in S(\mathbf{w}_{ij})}dt_\nu e^{-W^\out_\nu  t_v}\delta(\tau-\sum_\nu t_\nu)$  is the probability of completing the walk $\mathbf{w}_{ij}$ in a time $\tau$, see  \cite{crooks1998nonequilibrium,peliti2021stochastic,esposito2010three,seifert2012stochastic}. Here, $S(\mathbf{w}_{ij})$ is the set of states along the walk $\mathbf{w}_{ij}$; $W^\out_\nu=\sum_{j\neq\nu} W_{j\nu}$ is the total out rate from state $\nu$; and $t_\nu$ is the time spent in state $\nu$.
\begin{figure}[!hbt]
    \centering
    \includegraphics[width=1\columnwidth]{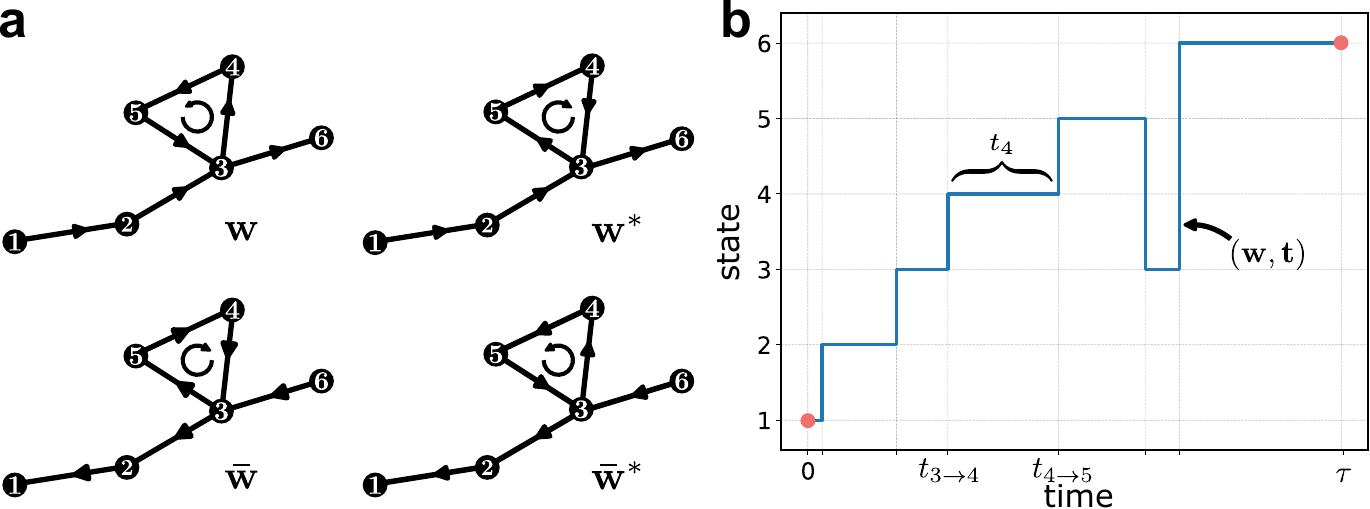}
    \caption{(a) A walk from state 1 to state 6, and the corresponding reversed walk $\mathbf{\bar{w}}$, partial-reversal operation $\mathbf{w}^*$ and the joint-reversed walk $\mathbf{\bar{w}}^*$. (b) An example trajectory along the walk $\mathbf{w}$ with a series of waiting times on all states during a time lag $\tau$.}\label{fig:trajectory}
\end{figure}

We introduce reversal and partial reversal operations on the trajectories. The reversal operation $\bar{\mathbf{w}}$  simply reverses the orientation of all edges in the trajectory, see Fig.~\ref{fig:trajectory}a.
In contrast, the partial-reversed trajectory $\mathbf{w}^*$ reverses the direction of all cycles in the trajectory, leaving the direction of edges that do not belong to cycles unaltered.
These two operations do not affect the times spent in each state along the walk, so that $P_{\mathbf{w}_{ij}}^\tau=P_{\mathbf{w}_{ij}^*}^\tau=P_{\mathbf{\bar{w}}_{ij}}^\tau=P_{\mathbf{\bar{w}}_{ij}^*}^\tau$. A walk and its partial-reversed counterpart both belong to the set of all possible walks with the same start and end. This means that a partial-reversal operation is a bijection in this set. In contrast, the reversal operation is a bijection from a set of trajectories to a set with swapped start and end states. Thus we have:
 \begin{equation}
\sum_{\mathbf{w}_{ji}}P(\mathbf{w}_{ji};\tau)=
\sum_{\mathbf{\bar{w}}_{ij}}P(\mathbf{\bar{w}}_{ij};\tau)=\sum_{\mathbf{\bar{w}}_{ij}^{*}}P(\mathbf{\bar{w}}_{ij}^*;\tau).
\end{equation}
Given these properties, we now obtain the bound on the relative propagator:
\begin{equation}\label{eq:boundproof}
\begin{aligned}
R_{i|j}^\tau&= \frac{\sum_{\mathbf{w}_{ij}}P(\mathbf{w}_{ij};\tau)}{\sum_{\mathbf{{w} }_{ji}}P(\mathbf{w}_{ji};\tau)}   
    = \frac{\sum_{\mathbf{w}_{ij}}P(\mathbf{w}_{ij};\tau)}{\sum_{\mathbf{\bar{w}^*}_{ij}}P(\mathbf{\bar{w}}^*_{ij};\tau)}\\
&=\frac{\sum_{\mathbf{w}_{ij}}\left(P_{\mathbf{w}_{ij}}^\tau\prod_{e\in\mathbf{w}_{ij}}W_e^+\right)}{\sum_{\mathbf{\bar{w}}_{ij}^*}\left({P_{\mathbf{\bar{w}}^*_{ij}}^\tau\prod_{e\in\mathbf{\bar{w}^*}_{ij}}W_e^+}\right)}\\
    &\leq \max_{\mathbf{w}_{ij}}\frac{P_{\mathbf{w}_{ij}}^\tau\prod_{e\in\mathbf{w}_{ij}}W_e^+}{P_{\mathbf{\bar{w}}^*_{ij}}^\tau\prod_{e\in\mathbf{\bar{w}^*}_{ij}}W_e^+}\\
    &=\max_{\mathbf{r}_{ij}}e^{s_{\mathbf{r}_{ij}}^{\res}},
\end{aligned}
\end{equation}
where in going from the second to the third line we used that
\begin{equation}\label{eq:inequlity}
\frac{\sum_i y_i}{\sum_i x_i}=\frac{\sum_i x_i( y_i/x_i)}{\sum_i x_i} \leq \max_i \frac{y_i}{x_i} \text{ if } x_i,y_i >0 \ \forall i.
\end{equation}
We note that the cycles in $\mathbf{w}_{ij}$ and $\mathbf{\bar{w}}_{ij}^*$ share the same orientation, as exemplified in Fig.~\ref{fig:trajectory}a. So when we evaluate $\prod_{e\in\mathbf{w}_{ij}}W_e^+/\prod_{e\in\mathbf{\bar{w}^*}_{ij}}W_e^+$, the contributions from cycles cancel out and only the remaining path contributes. The lower bound can be obtained following the same logic. This completes the proof of Eq.~\eqref{eq:conditional}.

{\em First affinity bound.} The upper and lower bounds in Eq.~\eqref{eq:conditional} do not depend on $\tau$. By choosing two arbitrary time lags $\tau$ and $\tau'$, we immediately obtain our first affinity bound
\begin{equation}\label{eq:firstaffinity}
\frac{R_{i|j}^{\tau}}{R_{i|j}^{\tau'}}\leq \frac{e^{s_{\mathbf{r}_{ij}}^{\res,\max}}}{e^{s_{\mathbf{r}_{ij}}^{\res,\min}}}\leq e^{\mathcal{F}_{{\mathbf{c}}}^\mathrm{sum}}.
\end{equation}
where $\mathcal{F}_{{\mathbf{c}}}^{\mathrm{sum}}=\sum_{\mathbf{c}} |\mathcal{F}_\mathbf{c}|$ is the sum of affinities of all cycles in the network. The second inequality can be proven as follows. Since $s^\res_{\mathbf{r}_{ij}}$ is antisymmetric with respect to changing orientation of the path, we have $s_{\mathbf{r}_{ij}}^{\res,\min}=-s_{\mathbf{r}_{ji}}^{\res,\max}$. The combination of the maximum dissipative paths from $j$ to $i$ and $i$ to $j$ form a closed walk. The contribution to the dissipation in this walk comes from the enclosed cycles. Since the closed walk is formed by two paths, any enclosed cycle can only be passed once. Therefore, the upper bound is smaller than the sum of all cycle affinities. In principle, the bound in Eq.~\eqref{eq:firstaffinity} can be made tightest by choosing $\tau$ and $\tau'$ in such a way to maximize the numerator and minimize the denominator on the left-hand side, respectively.

{\em Affinity bound on the time-reversal asymmetry.} We now seek for a bound for the relative joint probability $R_{i;j}^{\tau}\equiv P_{i;j}^\tau /P_{i;j}^\tau$. To this aim, we exploit the matrix-tree theorem, that bounds the steady-state probabilities as 
 \begin{equation}\label{eq:prob_bound}e^{s_{\mathbf{r}_{ji}}^{\res,\min}} \le \frac{p_{j}^{\sst}}{p_i^{\sst}} \le e^{s_{\mathbf{r}_{ji}}^{\res,\max}} ,
 \end{equation}
see \cite{maes2013heat,liang2022universal}. We recover this bound as a particular case of Eq.~\eqref{eq:conditional} for $\tau\rightarrow \infty$. We combine Eq.~\eqref{eq:conditional} and Eq.~\eqref{eq:prob_bound} to find $\ln R_{i;j}^{\tau}\in[s_{\mathbf{r}_{ij}}^{\res,\min}+s_{\mathbf{r}_{ji}}^{\res,\min},s_{\mathbf{r}_{ij}}^{\res,\max}+s_{\mathbf{r}_{ji}}^{\res,\max}]$, and therefore 
\begin{equation}\label{eq:boundjoint1}
    \mathcal{A}_{i;j}^{\tau}\leq s_{\mathbf{r}_{ij}}^{\res,\max}+s_{\mathbf{r}_{ji}}^{\res,\max}\leq \mathcal{F}_{{\mathbf{c}}}^\mathrm{sum}.
\end{equation}

An alternative bound for the relative joint probability can be obtained by considering Eq.~\eqref{eq:conditional} and multiplying each term by the positive quantity $p_j^\sst/p_i^\sst$. Defining the total entropy production rate $s_{\mathbf{r}_{ij}}^{\tot}=s_{\mathbf{r}_{ij}}^{\res}+\Delta s^{\sys}=s_{\mathbf{r}_{ij}}^{\res}-\ln p_i^\sst+\ln p_j^\sst$, we obtain
\begin{equation}\label{eq:boundjoint2}
 e^{s_{\mathbf{r}_{ij}}^{\tot,\min}}\leq R_{i;j}^{\tau} \leq e^{s_{\mathbf{r}_{ij}}^{\tot,\max}},
\end{equation}
where $s_{\mathbf{r}_{ij}}^{\tot,\min/\max}$ is the minimum/maximum total entropy production among all possible paths $\mathbf{r}_{ij}$.

\begin{figure}[!hbt]
    \centering
    \includegraphics[width=1\columnwidth]{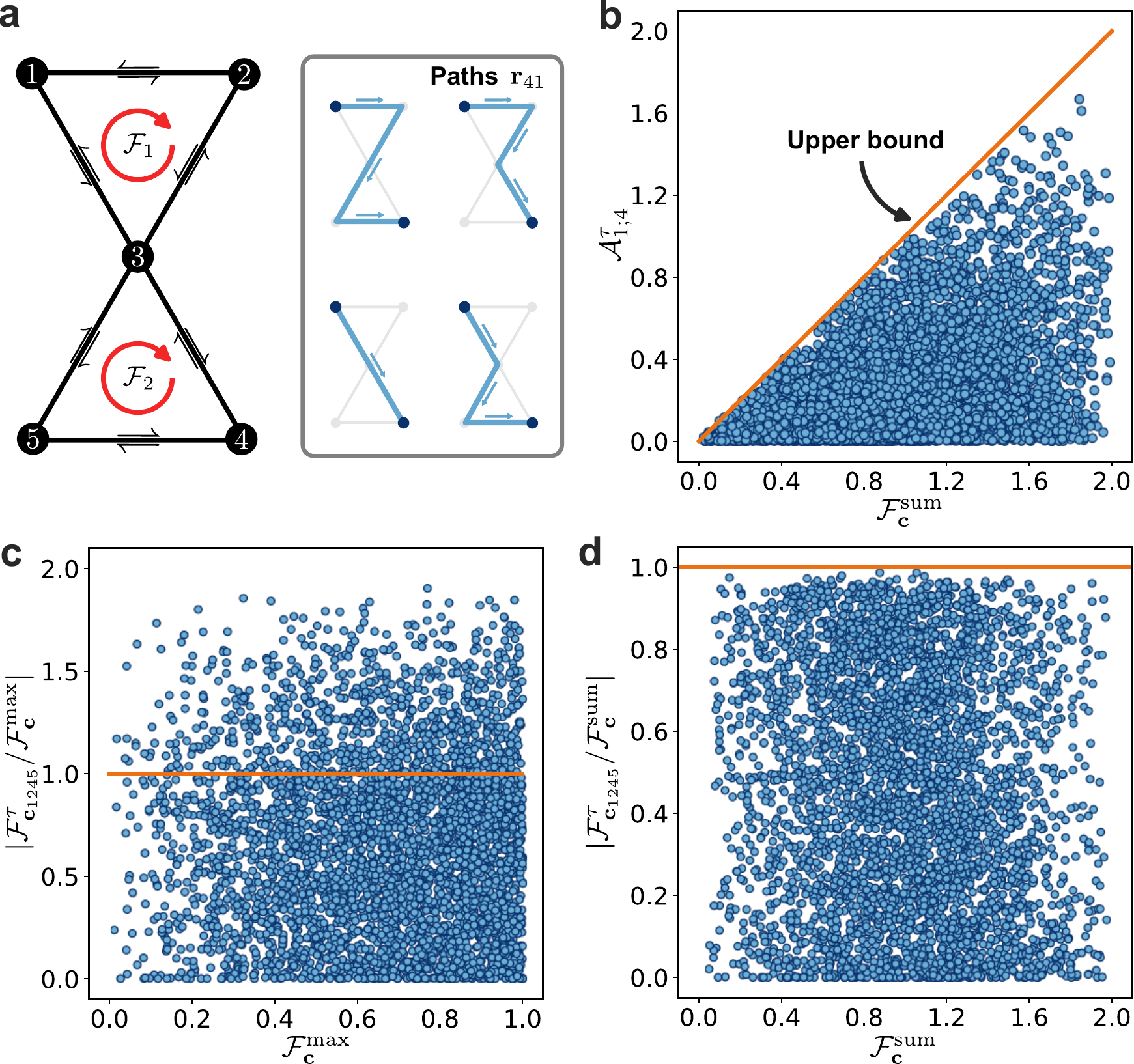}
    \caption{(a) A two-cycle network and the corresponding paths from state 1 to state 4. The network is brought out of equilibrium by two cycle affinities. (b) The thermodynamic bound on time-reversal asymmetry $\mathcal{A}_{1;4}^\tau$. The sum of cycle affinities $\mathcal{F}_{{\mathbf{c}}}^{\mathrm{sum}}=|\mathcal{F}_1|+|\mathcal{F}_2|$ sets an upper bound on the time-reversal asymmetry. (c) The bound on finite-time cycle affinity of the cycle $\mathbf{c} = (e_{12},e_{24},e_{45},e_{51})$ can exceed the maximum cycle affinity evaluated from the original transition matrix. (d) The finite-time cycle affinity is bounded from above by the sum of cycle affinities. All data points are obtained with $\tau=1$ and randomly generated transition rates such that $\mathcal{F}_1,\mathcal{F}_2\in[-1,1]$.}\label{fig:joint_asymmetry}
\end{figure}

{\em Asymmetry of directed fluxes.} In the short time limit, the leading term of the propagator between two connected states is $P_{i|j}^{\tau}\simeq W_{ij}\tau+\mathcal{O}(\tau^2)$. Therefore, the joint probability in a short lag time is related to the instantaneous directed flux:  $J_{j\to i}^{\sst}\equiv W_{ij}p_j^{\sst} =\lim_{\tau\rightarrow 0} P_{i;j}^\tau/\tau $. We obtain a bound on the asymmetry of directed fluxes in terms of the maximum cycle affinity:
\begin{equation}\label{eq:boundflux}
\begin{aligned}
\lim_{\tau\to 0}\mathcal{A}_{i;j}^\tau&=\left|\ln \frac{J_{j\to i}^\sst}{J_{i\to j}^\sst}\right|
=\left|\ln\frac{W_{ij}p_j^{\sst}}{W_{ji}p_i^{\sst}}\right|\\
&\leq |s_{e_{ij}}^{\res}+s_{\mathbf{r}_{ji}}^{\res,\max}|
= \mathcal{F}_{\mathbf{c}_{ij}}^{\max}.
\end{aligned}
\end{equation}
where $\mathcal{F}_{\mathbf{c}_{ij}}^{\max}$ is the cycle affinity maximized over all cycles that contain the edge from state $j$ to state $i$. We prove Eq.~\eqref{eq:boundflux} by combining the LDB condition, Eq.~\eqref{eq:ldb}, and the bound on steady-state probabilities, Eq.~\eqref{eq:prob_bound}. The combination of the edge from $j$ to $i$ and the maximum dissipative path from $i$ to $j$ forms at most one cycle. Therefore, the upper bound is given by a maximum cycle affinity, resulting in a tighter bound than the finite-time one given by Eq.~\eqref{eq:boundjoint1}.  

{\em Asymmetry of cross-correlations.} The correlation of two observables $a=(a_1,\dots,a_n)$ and $b=(b_1,\dots,b_n)$, with lag time $\tau$ is a useful measure of the non-equilibrium behavior of a system \cite{qian2004fluorescence}. At steady state, this correlation is defined by:
\begin{equation}
C_{ab}^{\tau}\equiv  \langle a(t+\tau)b(t)\rangle = \sum_{i,j}a_i P_{i;j}^\tau b_j.
\end{equation}
We first link the relative cross-correlation with the relative joint probability:
\begin{equation}\label{eq:corr}
\begin{aligned}
\frac{C_{ab}^\tau}{C_{ba}^\tau}
=\frac{\sum_{i,j}a_iP_{i;j}^\tau b_j}{\sum_{i,j}a_jP_{i;j}^\tau b_i}
=\frac{\sum_{i,j}a_iP_{i;j}^\tau b_j}{\sum_{i,j}a_iP_{j;i}^\tau b_j}
\leq \max_{i,j}R_{i;j}^\tau \le e^{\mathcal{F}_{\bf c}^{\rm sum}},
\end{aligned}
\end{equation}
which holds for physical observables such that the components of $a$ and $b$ are positive. The two inequalities in Eq.~\eqref{eq:corr} follow from Eq.~\eqref{eq:inequlity} and Eq.~\eqref{eq:boundjoint1}. 

One direct application of Eq.~\eqref{eq:corr} is a bound on the time-reversal asymmetry between coarse-grained states. We define $P_{I|J}(\tau) = \sum_{i\in I}\sum_{j\in J}P_{i|j}(\tau)[p_j^{st}/p_I^{\sst}]$, where $I$ and $J$ are given sets of states, and the weight $p_i^{\sst}/p_{I}^{\sst}$ is due to the lack of information on the starting state in $J$. We immediately obtain
\begin{equation}
\mathcal{A}_{I;J}^{\tau}\equiv\left|\ln\frac{P_{I|J}(\tau)p_J^{\sst}}{P_{J|I}(\tau)p_I^{\sst}}\right|
\!=\! \left|\ln \frac{\sum_{i\in I,j\in J}P_{i|j}(\tau)p_j^{\sst}}{\sum_{i\in I,j\in J}P_{j|i}(\tau)p_i^{\sst}}\right|\leq \mathcal{F}_{\bf c}^{\rm sum} ,
\end{equation}
where the last inequality is obtained from Eq.~\eqref{eq:corr} by choosing the indicator functions of the sets $I$ and $J$ as the functions $a$ and $b$.

We further obtain an affinity bound on the asymmetry of cross-correlations by a transformation of Eq.~\eqref{eq:corr}:
\begin{equation}
 \chi_{ab}\equiv\left|\frac{C_{ab}^{\tau}-C_{ba}^{\tau}}{C_{ab}^{\tau}+C_{ba}^{\tau}}\right|\leq \tanh[\mathcal{F}_{{\mathbf{c}}}^\mathrm{sum}/2].
\end{equation}

{\em Temporal-coarse-grained Markov process.}
A continuous-time master-equation can be coarse-grained in time to a Markov chain with time step $\tau$, $\Delta \mathbf{p} = \tau W^\tau \mathbf{p}$, by rearranging Eq.~\eqref{eq:finite_time_evol}. Here $W^{\tau} \equiv (e^{W\tau}-I)/\tau$ is a temporal-coarse-grained stochastic matrix which reduces back the original matrix $W$ in the $\tau\to 0$ limit. We note that the resulting Markov chain is fully connected. One natural question is how the thermodynamic properties are affected by coarse-graining \cite{puglisi2010entropy,chiuchiu2018mapping,cisneros2023dissipative}.  It was recently suggested that the maximum cycle affinity evaluated over cycles of a uniform cycle decomposition for the coarse-grained stochastic matrix $W^{\tau}$ might be smaller or equal to the maximum cycle affinity associated with the original stochastic matrix $W$ \cite{van2023dissipation}. We have simulated a non-equilibrium model on a butterfly network, and have found that the maximum cycle affinity bound does not hold for an arbitrary coarse-grained cycle, see Fig.~\ref{fig:joint_asymmetry}c. However, with our formalism, we find that the finite-time cycle affinity $\mathcal{F}_{{\mathbf{c}}}^{\tau}$ is bounded by the sum of the cycle affinities of the original network:
\begin{equation}\label{eq:Fctau_bound}
    \mathcal{F}_{{\mathbf{c}}}^{\tau}\equiv \ln\left(\prod_{e\in \mathbf{c}}\frac{W^{\tau+}_{e}}{W^{\tau-}_{e}}\right)\leq\mathcal{F}_{{\mathbf{c}}}^{\mathrm{sum}},
\end{equation}
see Fig.~\ref{fig:joint_asymmetry}d. This bound agrees with the conjecture in Ref.~\cite{van2023dissipation} in the unicycle case. To prove Eq.~\eqref{eq:Fctau_bound}, we pick two arbitrary states $k$ and $l$ from cycle $\mathbf{c}$ and decompose the cycle into two distinct paths between the two states, that we denoted by $\mathbf{r}_{kl}^\mathbf{c}$ and $\mathbf{r}_{lk}^\mathbf{c}$. We then split the finite-time cycle affinity into two temporal coarse-grained paths:
\begin{equation}\label{eq:twopaths}
    \mathcal{F}_{{\mathbf{c}}}^{\tau} = \ln\left[ \prod_{e\in \mathbf{r}_{kl}^\mathbf{c}}\frac{W^{\tau+}_{e}}{W^{\tau-}_{e}} \right]+\ln\left[ \prod_{e\in \mathbf{r}_{lk}^\mathbf{c}} \frac{W^{\tau+}_{e}}{W^{\tau-}_{e}}\right] \, .
\end{equation}
We can decompose the products of temporal-coarse-grained transition rates in Eq.~\eqref{eq:twopaths} into walks, similarly to how we have done in Eq.~\eqref{eq:walktau}. The resulting expression would contain the same product of the original rates as in Eq.~\eqref{eq:walktau}. The main difference is that, in this case, the temporal factor analogous to $P^\tau_{\mathbf{w}_{ij}}$ represents 
the probability of passing through the states composing the walk after each time lag $\tau$. In any case, as for the time-continuous case, this temporal factor is invariant under reversal and partial reversal operations, and thus cancels out in the final expression. We therefore obtain
\begin{equation}
 \ln\left[ \prod_{e\in \mathbf{r}_{kl}}\frac{W^{\tau+}_{e}}{W^{\tau-}_{e}} \right] \in [s_{\mathbf{r}_{kl}}^{\res,\min}, s_{\mathbf{r}_{kl}}^{\res,\max}],
\end{equation}
and also the same for the $\mathbf{r}_{lk}^\mathbf{c}$. By combining the bounds for the two paths, we finally reach Eq.~\eqref{eq:Fctau_bound}.

{\em Discussion.} In this Letter, we have introduced the measure of time-reversal asymmetry $\mathcal{A}^\tau_{i;j}$. Our main result is that $\mathcal{A}^\tau_{i;j}$ is bounded by the sum of cycle affinities in the system. Our bound directly connects the time-reversal asymmetry with its causes. While the temporal asymmetry depends on an adjustable time scale $\tau$ and the chosen pair of states, our bound solely depends on the cycle affinities.  Our main results lead to a bound on the cross-correlation asymmetry, which complements recent results on this subject \cite{ohga2023thermodynamic,van2023dissipation,shiraishi2023entropy}. In the short-time limit, $\mathcal{A}^\tau_{i;j}$ provides information on the asymmetry of directed fluxes. 

Further work should explore other observable quantifying temporal asymmetry that can be bounded using this approach. Additionally, this approach should be extended to continuous space stochastic processes described by Langevin equations and to complex chemical reaction networks \cite{ge2016nonequilibrium,peng2020universal,rao2016nonequilibrium}. The latter study could provide useful insight into biochemical systems.

\begin{acknowledgments}
We thank Ruicheng Bao, Yuansheng Cao, and Tan Van Vu for comments on a preliminary version of this manuscript. S.L. was supported by Swiss National Science Foundation under grant 200020\_178763 and JSPS Strategic Fellowship No. GR22LO6.  
\end{acknowledgments}

\bibliography{refs.bib}

\end{document}